\documentclass[11pt]{article}
\usepackage[margin=0.88in]{geometry}
\usepackage{graphicx}
\usepackage{booktabs}
\usepackage{amsmath}
\usepackage{amssymb}
\usepackage{microtype}
\usepackage{hyperref}
\usepackage{enumitem}
\usepackage{caption}
\usepackage{xcolor}
\usepackage{natbib}
\usepackage{authblk}
\usepackage{array}
\usepackage{multirow}
\usepackage{longtable}
\usepackage{url}
\usepackage{float}
\hypersetup{colorlinks=true,linkcolor=blue,citecolor=blue,urlcolor=blue}
\title{From Prompt to Recommendation:\\
A Fitted Stage Model of Brand Visibility in AI Search}
\author[1]{Benjamin Tannenbaum\thanks{Correspondence: ben@getaiso.com}}
\affil[1]{Aiso Boost Ltd.}
\date{September 2026}

\begin{document}
\maketitle

\begin{abstract}
What makes a brand appear in an AI answer? Page relevance alone cannot answer the question because generative search inserts engine-mediated decisions between the request and the recommendation. We analyze 34,960 unbranded prompt-engine observations from 75 anonymized Aiso projects, covering 2,854 distinct monitored prompts and repeated GPT and Gemini runs from June--September 2026. When neither the target brand nor its own domain appears in the observable live retrieval path, target mention rates are 2.8\% for GPT and 3.8\% for Gemini. With an own-domain citation but no branded fan-out, they rise to 49.0\% and 58.4\%. When both own-domain exposure and a branded fan-out occur, mention rates reach 91.4\% and 100\%.

The relationship persists within the same project, prompt, and engine across repeated runs: among prompt cells that vary in own-domain exposure while holding branded fan-out absent, exposure is associated with a mean mention-rate increase of 40.2 percentage points on GPT and 49.0 points on Gemini. Prior visibility is independently persistent. A previous non-mention plus no current own-domain exposure yields next-run mention rates of 1.6\% and 1.9\%; previous mention plus current exposure yields 80.5\% and 83.7\%.

We fit a chronological diagnostic model using prior-run history and contemporaneous retrieval indicators:
\[
\operatorname{logit}P(M_t=1)=\alpha_e+\beta_e\operatorname{logit}(\widetilde P_{t-1})+\gamma_e E_t+\delta_e F_t+\theta_e^\top X.
\]
On the latest 30\% holdout, the full model achieves AUC 0.963 on GPT and 0.942 on Gemini, compared with 0.937/0.917 for prior history alone and 0.880/0.840 for live signals alone. A manually curated prompt sensitivity gives nearly identical AUCs (0.960 and 0.943). A separate 199-prompt page-corpus validation finds that prompt-page match predicts Gemini exposure (AUC 0.641) more clearly than GPT exposure (0.545), placing relevance upstream of a larger engine-mediated exposure effect. The equation is predictive and observational, not a causal description of proprietary engine internals.
\end{abstract}

\section{Introduction}

A marketer can ask a deceptively simple question: what makes a brand appear in an AI answer?

For traditional web search, practitioners often compress a much more complicated ranking system into a small number of intuitions. A page should be relevant to the query. The site should have authority, historically approximated through links, reputation, and other signals. Search-engine ranking is not literally "keywords plus backlinks," but the abstraction is useful because it separates a content-side relevance problem from a web-side authority problem \citep{brin1998,robertson2009}.

The same compression is tempting in generative search. A growing class of systems assigns pages an "AI-readiness," "GEO," or content score and implicitly treats that score as a proxy for the probability of being cited or recommended. The appeal is obvious. A deterministic page score is cheap, stable, and easy to explain. A live generative engine is expensive, rate-limited, personalized, non-stationary, and partly opaque.

Recent evidence makes the limits of that shortcut clearer. \citet{bajemon2026} show that query-agnostic page scoring has weak within-query citation signal in controlled experiments, while query-conditioned relevance is materially more informative. That result is important: a page is not simply "good for AI" in the abstract; it is good or bad relative to a request. Our companion study, \emph{Scoring With the Engine}, then examines the missing production layer and finds that identical prompts can produce almost disjoint citation sets across ChatGPT, Copilot, Google, and Perplexity \citep{tannenbaum2026engine}. If the page never enters the engine's exposed source set, a perfect page-side score cannot make it visible.

This paper asks what the practical factor model should therefore look like.

We use proprietary Aiso research datasets collected for different purposes but sharing a common measurement question. One dataset starts with real human prompts and records the search-query fan-outs produced by ChatGPT. A second connects prompt-level page fit to GPT and Gemini citations and brand mentions. A third repeats the same prompts ten times and separates direct web ranking, citations, and final brand mentions. Taken together, the datasets let us observe several links in the chain:

\begin{center}
\textbf{real demand $\rightarrow$ page match $\rightarrow$ search/fan-out $\rightarrow$ evidence exposure $\rightarrow$ brand selection.}
\end{center}

They also reveal a second path. A model can mention a brand without exposing or citing the brand's own domain at all. That path is consistent with training-time brand knowledge, prior model memory, third-party evidence not represented by an own-domain citation, or other unobserved information available to the model. We therefore retain a separate \emph{prior} term rather than forcing every brand mention through live retrieval.

Our compact mnemonic is

\begin{equation}
\boxed{\text{AI visibility} \;\approx\; \text{Match} \times \text{Exposure} \times \text{Selection} + \text{Prior}.}
\label{eq:mnemonic}
\end{equation}

Equation~\ref{eq:mnemonic} is deliberately not presented as a literal production algorithm. It is a measurement framework. The paper's central contribution is to show that each term corresponds to an observable failure mode and that their empirical strengths differ sharply.

We make six contributions.

\begin{enumerate}[leftmargin=*]
\item We define a stagewise model of generative-search visibility that separates real user demand, query-page fit, search/fan-out activation, evidence exposure, final answer selection, and citation-free prior behavior.
\item We analyze a clean subset of 90 numbered real-user prompts and show that search-query fan-out is highly intent-dependent: 78.3\% of commercial prompts trigger fan-out versus 3.6\% of informational prompts in this sample.
\item We construct a reproducible best-page match score over 275 pages and show that match has meaningful but engine-specific predictive value. It is informative for Gemini exposure and mention, weak for GPT, and far smaller than the observed exposure effect.
\item We quantify the exposure-selection link across 199 prompts. Own-domain citation is associated with roughly a tenfold increase in brand-mention probability on both GPT and Gemini.
\item We document a separate citation-free path using 160 repeated ChatGPT runs. Established competitors can appear in answers far more frequently than their own URLs are cited, including on prompts for which the system produces zero citations.
\item We connect these observations to our earlier measurement program on search density, request state, context, buyer response, and cross-engine source divergence \citep{tannenbaum2026arsd,tannenbaum2026prompt,tannenbaum2026context,tannenbaum2026purchase,tannenbaum2026engine}.
\end{enumerate}

The result is not a new single score. It is almost the opposite: a reason to resist single-score interpretations unless the score says which stage it measures.

\section{From Search Ranking to a Generative Visibility Pipeline}

\subsection{The traditional analogy}

Classic information retrieval separates, conceptually if not operationally, query-document relevance from broader authority and ranking signals. BM25 formalizes lexical relevance as a probabilistic term-weighting model \citep{robertson2009}. Link-based methods such as PageRank add a network-level signal about the importance or endorsement of documents \citep{brin1998}. Modern search ranking is far more complex than either component, but the separation remains useful: content can be relevant yet weakly authoritative, or authoritative yet poorly matched to the query.

Generative search inserts additional transformations between the human request and the final brand mention. The user prompt may not be the query sent to search. A system may decide not to search at all. It may rewrite one request into multiple queries. Retrieved pages may be reranked, truncated, or ignored. The final answer may mention a brand from model priors even when the brand's own page is absent from the cited set.

This means that the SEO intuition "relevance + authority" needs at least two more verbs: \emph{retrieve} and \emph{select}.

\subsection{A formal stage decomposition}

Let $R$ denote the active request state, $e$ an engine, and $t$ time. The request state may be larger than the last visible user message in a multi-turn conversation \citep{tannenbaum2026prompt,tannenbaum2026context}. Let:

\begin{itemize}[leftmargin=*]
\item $F=1$ if the engine activates an observable search/fan-out path for the request;
\item $E=1$ if brand-supporting evidence is exposed to the answer process;
\item $M=1$ if the target brand is mentioned or recommended in the final answer;
\item $Q$ be a continuous query-page match score;
\item $P$ denote the engine's citation-free prior propensity to mention the brand for the request.
\end{itemize}

By the law of total probability,

\begin{align}
P(M=1\mid R,e,t)
=&\; P(F=0\mid R,e,t)P(M=1\mid F=0,R,e,t) \nonumber\\
&+P(F=1\mid R,e,t)\{P(E=1\mid F=1,R,e,t) \\
&\quad P(M=1\mid E=1,F=1,R,e,t) \nonumber\\
&\quad+P(E=0\mid F=1,R,e,t)P(M=1\mid E=0,F=1,R,e,t)\}.
\label{eq:total}
\end{align}

Equation~\ref{eq:total} is an identity, not a behavioral assumption. It separates the no-search path from the search path and separates evidence exposure from final selection.

The page-match variable $Q$ is best understood as a predictor of exposure and selection, not as a probability that can simply be multiplied into Equation~\ref{eq:total}. We use the mnemonic in Equation~\ref{eq:mnemonic} because it maps cleanly to operational questions:

\begin{enumerate}[leftmargin=*]
\item \textbf{Match:} Is there a page that actually answers the real request?
\item \textbf{Exposure:} Does the engine's search, fan-out, retrieval, or citation process surface supporting evidence?
\item \textbf{Selection:} Given that evidence, does the model put the brand into the answer?
\item \textbf{Prior:} Can the brand appear without visible live evidence?
\end{enumerate}

\begin{figure}[H]
\centering
\includegraphics[width=0.98\linewidth]{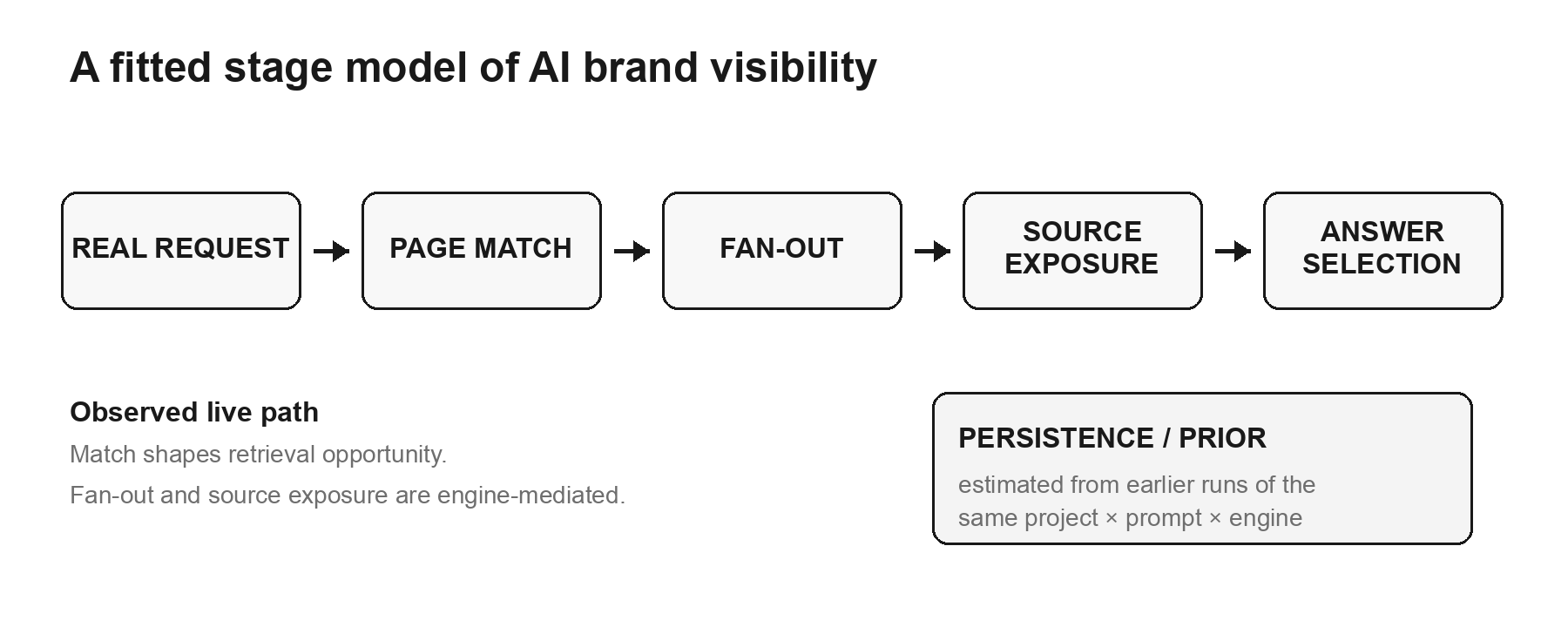}
\caption{The stage model used in this paper. The main live-retrieval path moves from real demand through page fit, search/fan-out, evidence exposure, and final selection. A separate prior-compatible path allows a brand to be mentioned without an observed citation.}
\label{fig:stage}
\end{figure}

Figure~\ref{fig:stage} summarizes the model.

\subsection{Why request state comes first}

The first four papers in this research program progressively changed the measurement unit. \citet{tannenbaum2026arsd} shows that one conversational answer can compress multiple traditional search actions and source inspections. \citet{tannenbaum2026prompt} shows that the final prompt often carries only a fraction of the observable request state accumulated across a conversation. \citet{tannenbaum2026context} demonstrates that restoring preceding conversation context materially changes answers. \citet{tannenbaum2026purchase} then shows that recommendation behavior is observable much more often than the buyer's eventual decision.

The implication for visibility measurement is straightforward. A page-match score against a synthetic standalone keyword panel can be statistically precise and still target the wrong demand distribution. We therefore begin this paper with real prompts and treat page fit as conditional on those prompts rather than as a property of the page alone.

\section{Related Work}

\subsection{Generative Engine Optimization}

The original GEO study formalized optimization for visibility in generative-engine responses and reported that several content interventions could increase visibility in controlled settings \citep{aggarwal2024}. Its importance was conceptual as much as numerical: it made the creator-side visibility problem measurable. The field has since broadened from isolated content interventions to questions of source mix, freshness, retrieval, competitive effects, and measurement design. \citet{martinez2026} surveys 45 studies from 2023--2026 and characterizes GEO as a stochastic, partially observable pipeline rather than a single ranking task. \citet{chen2025geo} similarly compares traditional and AI search across verticals, languages, and paraphrases, emphasizing that source behavior changes with both system and request formulation.

\citet{bajemon2026} revisits that program on modern engines. Their protocol is designed to validate a deterministic, manipulation-resistant content score without repeatedly querying an expensive oracle. The key result for the present paper is that generic page scoring has weak within-query association with citation order, while query-conditioned relevance is substantially more informative. Their fixed-candidate design is well suited to studying conditional citation preference. It does not estimate whether a live engine will expose a page organically.

Our companion study tests that missing layer and finds extremely low same-prompt citation overlap across four live AI engines \citep{tannenbaum2026engine}. That result motivates the engine-specific exposure term in Equation~\ref{eq:total}.

\subsection{Retrieval, RAG, and query rewriting}

Retrieval-augmented generation explicitly separates retrieval from generation \citep{lewis2020}. Self-RAG and related approaches make retrieval itself conditional, reinforcing the idea that search activation is a model decision rather than a constant \citep{asai2024}.

Conversational retrieval adds a second transformation: the user utterance may need rewriting before it is useful to a retriever. Explicit query rewriting and conversational dense-retrieval work show that the query actually used for retrieval can differ materially from the surface user message \citep{qian2022,wu2022,lin2021,ma2023rewrite,mo2023,ye2023}. Multi-query systems intentionally create several retrieval views of one request to increase document coverage \citep{rackauckas2024,li2024dmqr}. Yet higher recall does not guarantee final-answer inclusion: retrieval set selection, reranking, and context budgets can remove candidates later in the pipeline \citep{lee2025setr,medrano2026}. The observed ChatGPT fan-outs in our real-prompt cohort are a production analogue of this idea: one human request can become multiple search queries with a different intent mix.

\subsection{Live generative-search citations}

Recent audits treat citations themselves as an object of study rather than assuming a common source pool. \citet{yang2025} analyzes more than 366,000 citations from OpenAI, Perplexity, and Google systems and finds provider-specific source choices alongside shared concentration patterns. \citet{allaham2026} audits ChatGPT, Copilot, Gemini, and Perplexity on 712 real-world queries and reports both repeatedly cited domains and a long tail of minimally cited sources.

These studies align with the exposure view: citation is not simply a deterministic readout of page quality. It is a downstream consequence of engine-specific source selection. Large-scale audits reinforce the point. \citet{grossman2026} reports low source-set overlap across Google Search, AI Overviews, and Gemini, while \citet{strauss2025} distinguishes pages apparently visited from pages ultimately cited in search-enabled LLM conversations. Both results caution against treating displayed citations as a complete retrieval trace.

\subsection{Citation versus answer inclusion}

A page can be retrieved but not cited, cited but not substantively used, or absent from the visible citation list while the brand still appears in the answer. Work on citation attribution and RAG source influence therefore distinguishes retrieval, attribution, and generation outcomes \citep{gao2023,nematov2025}. Recent work also separates surface citation from actual source absorption or re-attribution after generation and compression \citep{mody2026}.

This distinction is central to our third cohort. Brands with strong model familiarity can be mentioned frequently even when their own domain is rarely cited. We use the neutral term \emph{prior-compatible path} because the data do not identify exactly whether the uncited mention comes from pretraining, prior model memory, third-party evidence, or another unobserved source.

\section{Data}

The paper combines three focused cohorts and a larger validation panel that identify different stages. Table~\ref{tab:datasets} makes those boundaries explicit.

\begin{table}[H]
\centering
\caption{Datasets and the stage each can identify. No single cohort is treated as a complete causal chain.}
\label{tab:datasets}
\small\setlength{\tabcolsep}{4pt}
\begin{tabular}{>{\raggedright\arraybackslash}p{0.19\linewidth}>{\raggedright\arraybackslash}p{0.22\linewidth}>{\raggedright\arraybackslash}p{0.24\linewidth}>{\raggedright\arraybackslash}p{0.25\linewidth}}
\toprule
Cohort & Observations & Directly observed & Primary use \\
\midrule
Real-prompt fan-out & 90 numbered rows, 80 unique real-user prompts; 42 fan-out queries from 20 prompts & Human prompt, prompt intent, ChatGPT search-query fan-out, fan-out intent & Search activation and request expansion \\
Organization A visibility & 199 prompts; GPT and Gemini source lists and brand mentions; 275-site-page crawl & Prompt, current page fit, own-domain citation, final brand mention & Match $\rightarrow$ exposure and exposure $\rightarrow$ selection \\
Organization B repeated runs & 16 prompts $\times$ 10 ChatGPT runs = 160; Bing rank and 13 Google fan-out subqueries & Direct rank, citation count, final brand mention, citation-free runs & Citation-free prior-compatible path and rank/citation/mention gaps \\
Large validation panel & 34,960 unbranded GPT/Gemini observations across 75 anonymous projects & Fan-out queries, source URLs, answer text, repeated history, intent & Prior + live-evidence model; repeated-run stratification \\
\bottomrule
\end{tabular}
\end{table}

\subsection{Cohort A: real-user prompts and ChatGPT fan-outs}

The first cohort comes from an Aiso conversation-intelligence workbook assembled in February 2026. The numbered source section contains 90 rows across legal/regtech, IT/technology, travel, and beauty. Exact duplicate prompt text is collapsed for the trigger analysis, leaving 80 unique user requests. The underlying prompts were extracted from real conversational-AI usage for research and commercial-intelligence purposes. Later researcher-added tests present elsewhere in the workbook are excluded from the analysis reported here.

For each prompt, the workbook records the original intent label and any observed ChatGPT \texttt{search\_query} fan-out. The clean subset contains 42 observed fan-out queries generated by 20 unique prompts. A triggered prompt produces a mean 2.10 fan-out queries.

We report descriptive rates and an approximate log-risk-ratio confidence interval. Because the sample was assembled for exploratory research rather than population inference, the confidence interval reflects sampling variation inside the observed cohort, not a claim that the cohort is representative of all ChatGPT usage.

\subsection{Cohort B: page match, citations, and brand mentions}

The second cohort is an Aiso AI-visibility export for Organization A containing 199 prompts and paired GPT/Gemini measurements. Each row contains the prompt, intent and funnel labels, whether Organization A was mentioned, an average answer position, the set of brands surfaced, and source URLs for GPT and Gemini.

We define own-domain evidence exposure as the presence of the target organization's registered domain URL in the engine's stored source list. This is intentionally conservative. A third-party page can mention Organization A and influence the answer even when Organization A's domain itself is absent. Therefore "no own-domain citation" does not mean "no Organization A-supporting evidence anywhere."

To measure page fit, we crawled the public Organization A site on 19 September 2026 using the WordPress/Yoast sitemaps. The crawl produced 275 usable post, page, and GEO URLs. Scripts, navigation, forms, and boilerplate were removed before scoring.

The visibility export is dated 6 July 2026, while the site crawl is later. The match analysis is therefore a retrospective coverage proxy, not a historical causal ranking experiment. Content added between July and September can raise the measured fit of prompts that were less well covered at the original run date. We return to this limitation later.

\subsection{Normalized BM25 best-page match}

We use a transparent lexical relevance score so the analysis does not depend on a proprietary embedding model. After lowercasing, tokenization, and a fixed stopword removal list, each site page is represented as a bag of terms. For prompt $q$ and page $d$, we compute

\begin{equation}
s(q,d)=
\frac{
\sum_{w\in q} IDF(w)
\frac{tf(w,d)(k_1+1)}
{tf(w,d)+k_1(1-b+b|d|/\overline{|d|})}
}{
(k_1+1)\sum_{w\in q}IDF(w)
},
\label{eq:bm25}
\end{equation}

with $k_1=1.5$ and $b=0.75$. The denominator uses the maximum BM25 term-saturation factor $(k_1+1)$, yielding an interpretable normalized score bounded approximately between 0 and 1 for the observed corpus. We define the prompt-level page match as

\begin{equation}
Q(q)=\max_{d\in \mathcal{D}_{A}} s(q,d).
\end{equation}

The median best-page match is 0.381, with first and third quartiles 0.291 and 0.494. We use continuous $Q$ for AUC analyses and quartiles for descriptive rates.

For uncertainty, we use 10,000 prompt-level bootstrap resamples. AUC is calculated as the probability that a randomly selected positive row has a higher match score than a randomly selected negative row, with ties receiving half credit.

\subsection{Cohort C: repeated ChatGPT runs and the citation-free path}

The third cohort is a May 2026 competitive AI-search benchmark for Organization B. Sixteen parent-decision prompts were each run ten times through \texttt{gpt-5.3-chat-latest} with web search and a United Kingdom user location, producing 160 answer runs. The workbook separately records brand mentions, citation surfaces, Bing top-30 direct-domain results, and a Google fan-out cross-tab.

The repeated-run design is useful because a single answer can exaggerate or hide stochastic variation. It also provides a direct observation of citation-free answers. Four of the sixteen prompts produced zero citations across all ten runs, yielding 40 answer runs in which visible citation evidence is absent.

We do not equate citation-free output with pure pretraining memory. Search could be unobserved, a tool may return uncited information, or the model may rely on latent knowledge. We therefore interpret these cases as evidence of a \emph{prior-compatible} path rather than proving a specific mechanism.

\section{Results}

\subsection{Real commercial prompts are far more likely to trigger fan-out}

Among the 80 unique real-user prompts, 20 generate at least one observed search-query fan-out. The overall trigger rate is 25.0\%.

The rate is highly intent-dependent. Eighteen of 23 commercial prompts trigger fan-out (78.3\%). Only 2 of 55 informational prompts do so (3.6\%). The risk ratio is 21.5, with an approximate 95\% interval of 5.4--85.3.

\begin{figure}[H]
\centering
\includegraphics[width=0.92\linewidth]{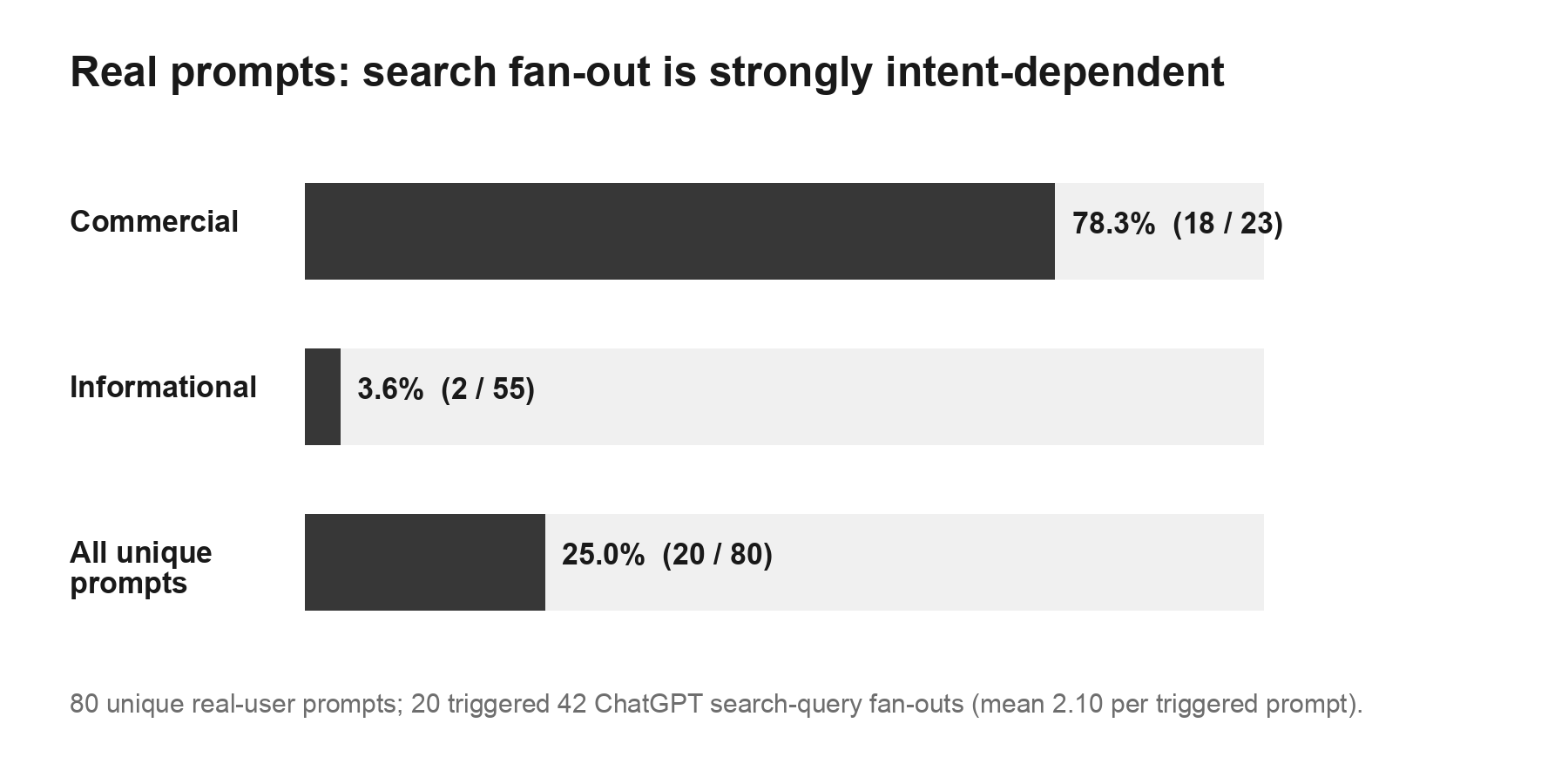}
\caption{Observed ChatGPT search-query fan-out activation in the clean real-user-prompt subset. Commercial requests trigger fan-out much more often than informational requests in this cohort.}
\label{fig:fanout}
\end{figure}

The 20 triggered prompts produce 42 search queries, a mean of 2.10 per prompt. Thirty-nine of the 42 fan-outs are labeled commercial, two branded, and one informational. Of the two informational source prompts that trigger fan-out, both move into a commercial search query. Of the 18 triggered commercial source prompts, 16 remain commercial, one becomes informational, and one becomes branded.

Table~\ref{tab:fanoutexamples} shows the mechanism with representative prompt-query pairs.

\begin{table}[H]
\centering
\caption{Examples of observed fan-out expansion. User prompts are shown as short excerpts.}
\label{tab:fanoutexamples}
\small\setlength{\tabcolsep}{4pt}
\begin{tabular}{>{\raggedright\arraybackslash}p{0.34\linewidth}>{\raggedright\arraybackslash}p{0.46\linewidth}>{\raggedright\arraybackslash}p{0.12\linewidth}}
\toprule
User-prompt excerpt & Observed search-query fan-out & Fan-out intent \\
\midrule
"...looking for a new mascara" & "what to look for in mascara how to choose mascara ingredients" & Info. \\
 & "best mascaras makeup recommendations" & Comm. \\
\addlinespace
"...suggest an accounting software (limit to 1)..." & "small business accounting software key features why choose one" & Comm. \\
 & "best accounting software for small business" & Comm. \\
\addlinespace
"...document management for law firms?" & "best open source document management software features DMS open source law firm use" & Comm. \\
 & "open source document management system for law firms free open source DMS" & Comm. \\
\bottomrule
\end{tabular}
\end{table}

The operational implication is not simply that commercial prompts search more. The engine frequently constructs an \emph{evaluation space}: best options, features, comparisons, ingredients, pricing, and selection criteria. The brand is competing against the fan-out representation, not only against the literal human wording.

\subsection{Page match is a real but engine-specific signal}

We next ask whether a better matching owned page makes source exposure more likely.

For GPT, best-page match is close to chance as a discriminator of own-domain citation: AUC 0.545 (95\% bootstrap CI 0.429--0.658). Its AUC for final brand mention is 0.539 (0.443--0.637). The top-versus-bottom match-quartile risk ratios are 1.33 for citation and 1.30 for mention; both intervals include one.

For Gemini, the signal is stronger. Match predicts own-domain citation with AUC 0.641 (0.556--0.722) and brand mention with AUC 0.608 (0.519--0.695). In the highest match quartile, Organization A's domain is cited on 50.0\% of prompts, compared with 22.0\% in the lowest quartile. The top-versus-bottom citation risk ratio is 2.27 (1.32--4.74). Brand mention rises from 22.0\% to 46.0\%, a risk ratio of 2.09 (1.20--4.29).

\begin{figure}[H]
\centering
\includegraphics[width=0.96\linewidth]{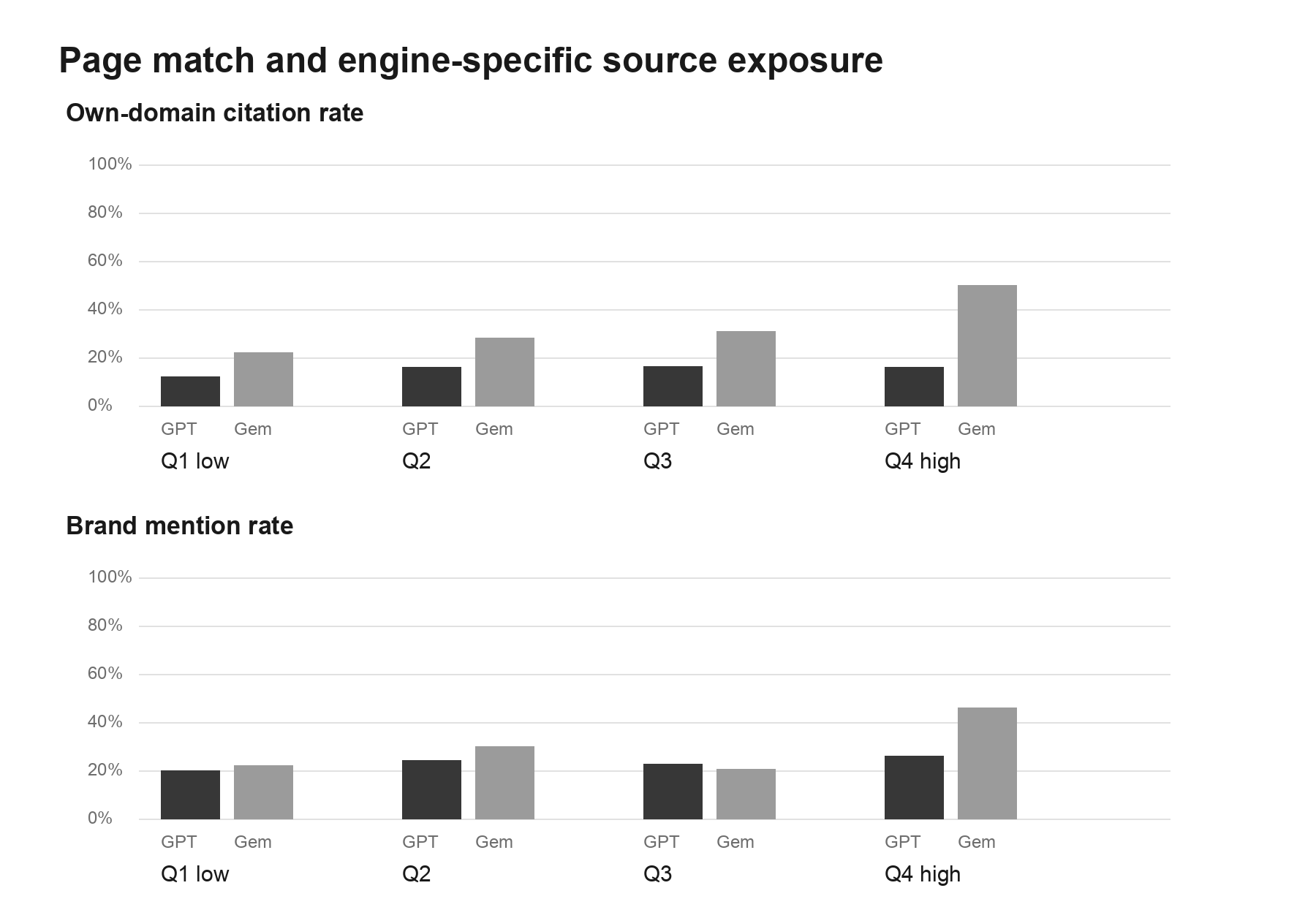}
\caption{Organization A best-page match quartiles versus own-domain citation and final brand mention. The match gradient is pronounced for Gemini and weak for GPT.}
\label{fig:match}
\end{figure}

Figure~\ref{fig:match} illustrates why a single engine-free relevance score should not be interpreted as a universal visibility score. The same match statistic has different downstream value by engine.

This finding is compatible with \citet{bajemon2026}: request-conditioned relevance contains more citation signal than generic page quality. Our result adds a production qualification. Even a useful relevance score can have different exposure consequences depending on the engine.

\subsection{Evidence exposure is associated with an order-of-magnitude selection effect}

The exposure-selection relationship is much larger.

GPT cites an Organization A-owned URL on 30 of the 199 prompts. Organization A is mentioned on all 30 of those rows. Among the 169 rows without an Organization A-domain citation, the brand is mentioned on 16. The observed mention rates are therefore 100\% versus 9.5\%, a risk ratio of 10.6 (95\% bootstrap CI 7.0--18.9).

Gemini cites an Organization A-owned URL on 65 prompts and mentions the brand on 48 of them, a 73.8\% selection rate conditional on own-domain citation. Among 134 rows without an own-domain citation, Organization A is mentioned 11 times (8.2\%). The risk ratio is 9.0 (5.5--18.9).

\begin{figure}[H]
\centering
\includegraphics[width=0.90\linewidth]{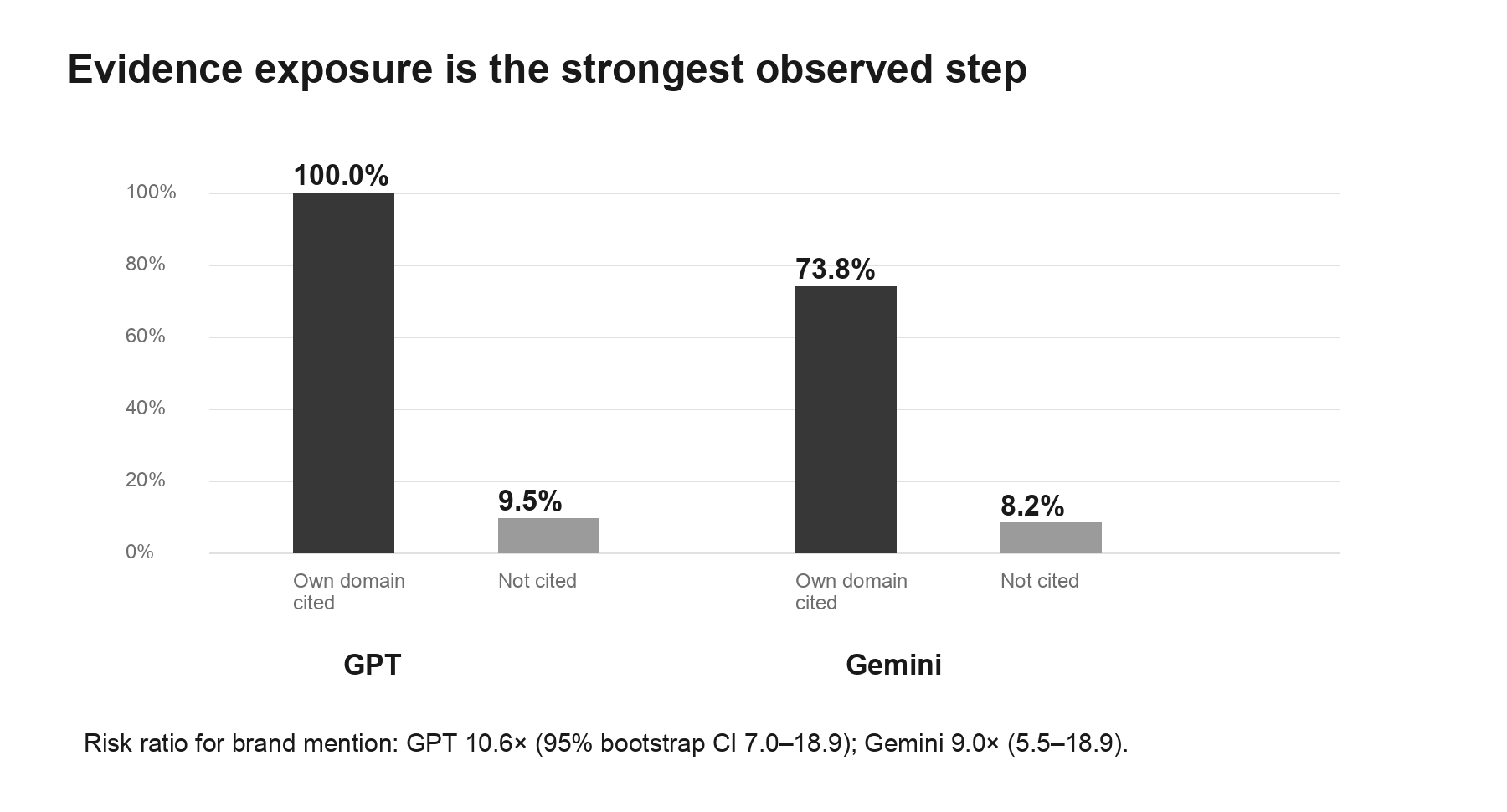}
\caption{Final brand-mention rate when the brand's own domain is present versus absent from the stored source list. This is an observational association, not a randomized causal effect.}
\label{fig:exposure}
\end{figure}

The magnitude matters. Page fit can double the Gemini citation rate between the bottom and top quartile. Actually appearing in the evidence set is associated with roughly a ninefold to elevenfold change in final brand mention.

This is the central empirical reason to separate Match from Exposure.

\subsection{High match without exposure still loses}

The highest match quartile contains 50 prompts. For Gemini, 25 of those 50 prompts cite Organization A's domain. Among those 25, the brand is mentioned on 21 (84\%). Among the 25 high-match prompts without own-domain citation, the brand is mentioned on only 2 (8\%).

For GPT, 8 of the 50 highest-match prompts cite Organization A, and all 8 mention it. Among the 42 high-match prompts without an own-domain citation, only 5 mention the brand (11.9\%).

In other words, high page fit is not a substitute for exposure in either engine. It is a favorable upstream condition whose downstream value depends on whether the engine surfaces evidence.

\subsection{The prior-compatible path is large for established competitors}

The Organization B benchmark shows why the visibility model also needs a path that does not require an own-domain citation.

Across 160 repeated ChatGPT runs:
\begin{itemize}[leftmargin=*]
\item Organization B is mentioned 12 times and its own URLs are cited 9 times.
\item Competitor 1 is mentioned 74 times and its own URLs are cited 7 times.
\item Competitor 2 is mentioned 65 times and its own URLs are cited 5 times.
\item Competitor 3 is mentioned 36 times and its own URLs are cited 11 times.
\end{itemize}

\begin{figure}[H]
\centering
\includegraphics[width=0.92\linewidth]{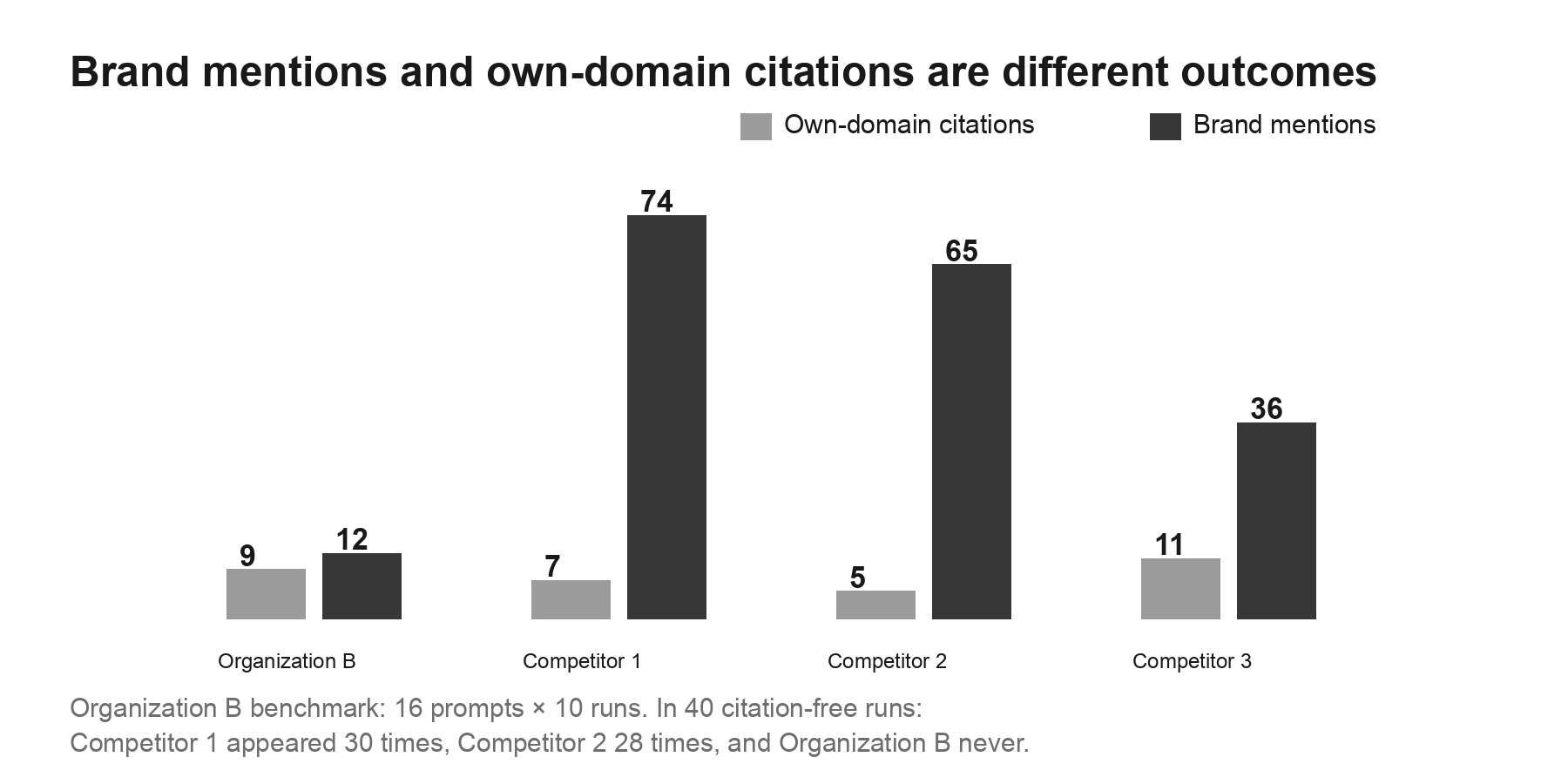}
\caption{Own-domain citation count versus brand-mention count across 160 repeated ChatGPT runs. Large mention-citation gaps indicate a substantial path not explained by direct citation of the brand's own domain.}
\label{fig:prior}
\end{figure}

The citation-free subset is even more informative. Four prompts produce zero citations in all ten repetitions, for 40 citation-free answer runs. Competitor 1 is mentioned in 30 of those 40 runs and Competitor 2 in 28. Organization B is mentioned in zero.

This pattern is compatible with a model-prior or brand-memory mechanism, but the data cannot prove that interpretation. What they do prove is a measurement fact: an own-domain citation model alone cannot account for final brand visibility.

The direct-rank comparison makes the same point from another angle. Organization B appears in Bing's top 30 for three of sixteen prompts. On one of them, the high-performance-athlete query, Organization B's dedicated page ranks in the direct search results but ChatGPT mentions Organization B in 0 of 10 runs. Competitor 3 appears in 9 of 10, Competitor 1 in 8 of 10, and Competitor 2 in 6 of 10. Direct rank creates an opportunity; it does not determine the composed answer.

\subsection{The effects are different in kind, not just size}

\begin{figure}[H]
\centering
\includegraphics[width=0.94\linewidth]{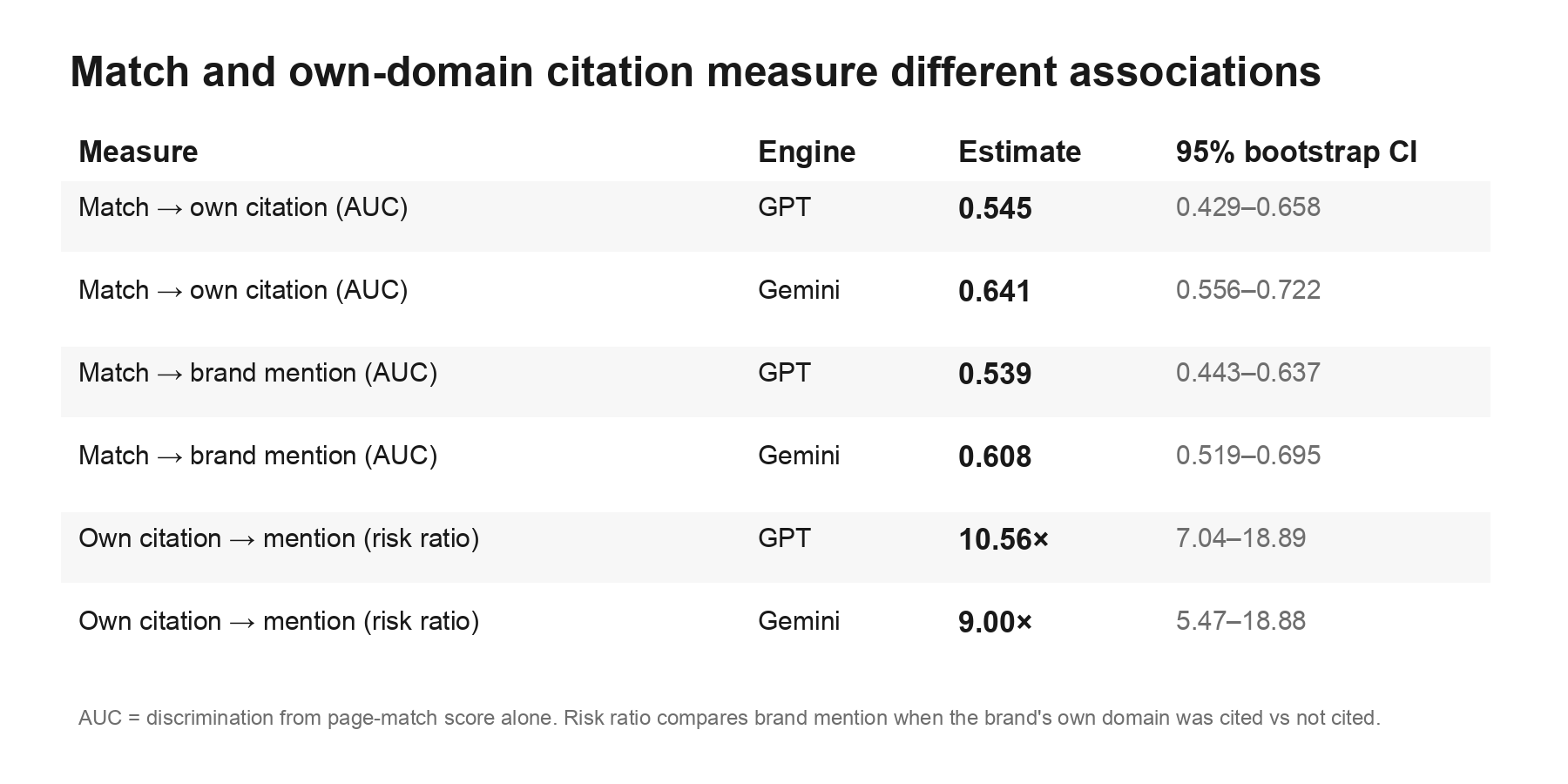}
\caption{Page match has modest, engine-specific discrimination. Own-domain evidence exposure has a much larger association with final brand selection.}
\label{fig:effects}
\end{figure}

Figure~\ref{fig:effects} summarizes the main estimates.

The distinction suggests a practical hierarchy:

\begin{enumerate}[leftmargin=*]
\item \textbf{Demand coverage} asks whether the brand has content for what people actually ask.
\item \textbf{Match quality} asks whether a specific page is aligned with the request or fan-out.
\item \textbf{Exposure} asks whether the engine actually surfaces brand-supporting evidence.
\item \textbf{Selection} asks whether the brand survives from evidence into the answer.
\item \textbf{Prior} captures visibility that occurs without an observed own-domain citation.
\end{enumerate}

A brand can fail at any stage.

\section{Fitted Stage Equation}

The earlier case analyses suggest a stage structure, but the longitudinal panel lets us estimate it directly. For repeated project-prompt-engine cells we define a smoothed historical prior using only earlier runs:
\[
\widetilde P_{t-1}=\frac{m_{<t}+0.1}{n_{<t}+2},
\]
where $m_{<t}$ is the number of earlier target mentions and $n_{<t}$ is the number of earlier observations. We then fit, separately by engine,
\begin{equation}
\operatorname{logit}P(M_t=1)=
\alpha_e+\beta_e\operatorname{logit}(\widetilde P_{t-1})
+\gamma_e E_t+\delta_e F_t+\theta_e^\top X_q,
\label{eq:stagefit}
\end{equation}
where $E_t$ is current own-domain source exposure, $F_t$ indicates that the engine generated a branded fan-out from an unbranded head prompt, and $X_q$ contains request-intent controls.

Equation~\ref{eq:stagefit} is the paper's fitted visibility equation. It is intentionally predictive rather than causal. $E_t$ and $F_t$ are engine-produced variables observed during the current answer process, not manipulable treatments. The model answers a diagnostic question: conditional on what was historically likely for this request, how much additional information is contained in the live retrieval path?

\section{Large-Panel Validation}

The case-study analyses above establish the stages qualitatively and on interpretable subsets. We next test the framework on the larger Aiso monitoring panel. To avoid a trivial branded-query effect, we exclude observations where the target organization's name appears in the user prompt. We also exclude observations generated after the start of this analysis. The resulting panel contains 34,960 GPT/Gemini engine-prompt-run observations spanning 75 projects; 73 projects have at least 20 unbranded observations per engine. Prompts come from Aiso's proprietary enrichment and monitoring pipeline. Organization identities are not reported.

For each engine run we derive two live-evidence indicators. \emph{Own-domain exposure} equals one when at least one stored source URL belongs to the target organization's domain. \emph{Branded fan-out} equals one when the target organization's name occurs in at least one observed search-query fan-out. The outcome is literal target-brand inclusion in the answer, cross-checked against Aiso's brand extraction tables where available. These indicators are post-retrieval diagnostics; they should not be used as pre-run causal predictors.

\subsection{A four-cell live-evidence ladder}

Table~\ref{tab:ladder} gives the simplest result. With neither live signal present, target-brand mention is rare: 2.8\% on GPT and 3.8\% on Gemini. Own-domain exposure alone raises the rates to 49.0\% and 58.4\%. A branded fan-out without an own-domain citation is uncommon but highly predictive, at 64.4\% and 84.8\%. When both are present, mention rates are 91.4\% on GPT and 100\% on Gemini.

\begin{table}[H]
\centering
\caption{Brand mention by observed live-evidence state, unbranded prompts only.}
\label{tab:ladder}
\begin{tabular}{lrrr}
\toprule
Engine and evidence state & $n$ & Mentions & Rate \\
\midrule
GPT: neither & 15,524 & 438 & 2.8\% \\
GPT: own domain only & 1,769 & 866 & 49.0\% \\
GPT: branded fan-out only & 59 & 38 & 64.4\% \\
GPT: both & 128 & 117 & 91.4\% \\
\addlinespace
Gemini: neither & 13,801 & 524 & 3.8\% \\
Gemini: own domain only & 3,415 & 1,995 & 58.4\% \\
Gemini: branded fan-out only & 33 & 28 & 84.8\% \\
Gemini: both & 231 & 231 & 100.0\% \\
\bottomrule
\end{tabular}
\end{table}

\begin{figure}[H]
\centering
\includegraphics[width=0.94\linewidth]{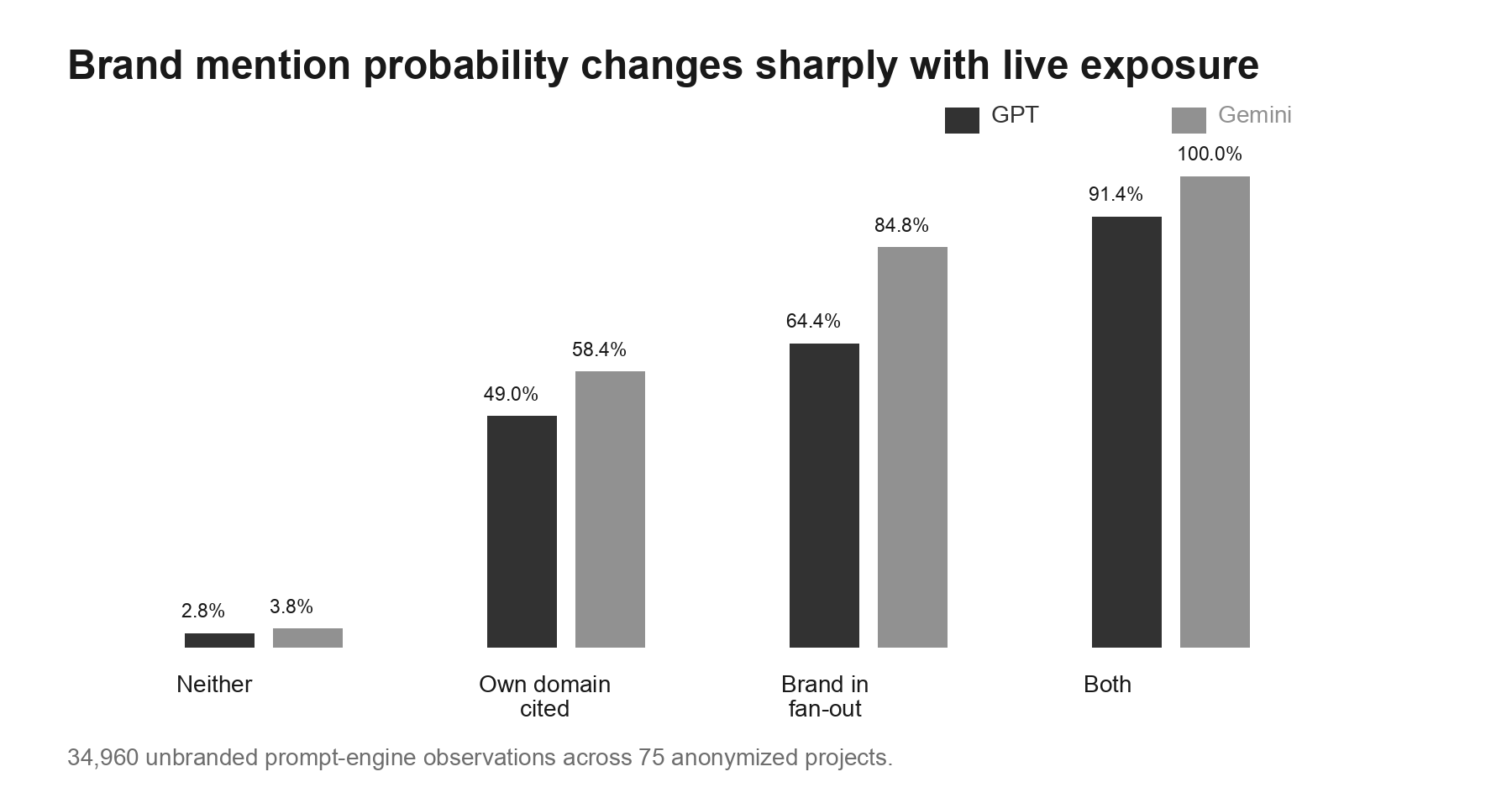}
\caption{Observed brand-mention rates by live-evidence state in the large unbranded-prompt panel.}
\label{fig:ladder}
\end{figure}

Relative to the neither-signal baseline, own-domain exposure alone corresponds to a 17.4-fold GPT and 15.4-fold Gemini increase in mention probability. The combination of own-domain exposure and branded fan-out corresponds to 32.4-fold and 26.3-fold increases. These are descriptive risk ratios, not causal multipliers.

\subsection{Within-prompt repeated-run evidence}

Stable brand quality and prompt difficulty can confound the cross-sectional ladder. We therefore exploit repeated observations of the same organization-prompt-engine cell. There are 1,837 such cells with at least two runs per engine. Own-domain exposure varies within 499 GPT cells and 558 Gemini cells.

\begin{figure}[H]
\centering
\includegraphics[width=0.90\linewidth]{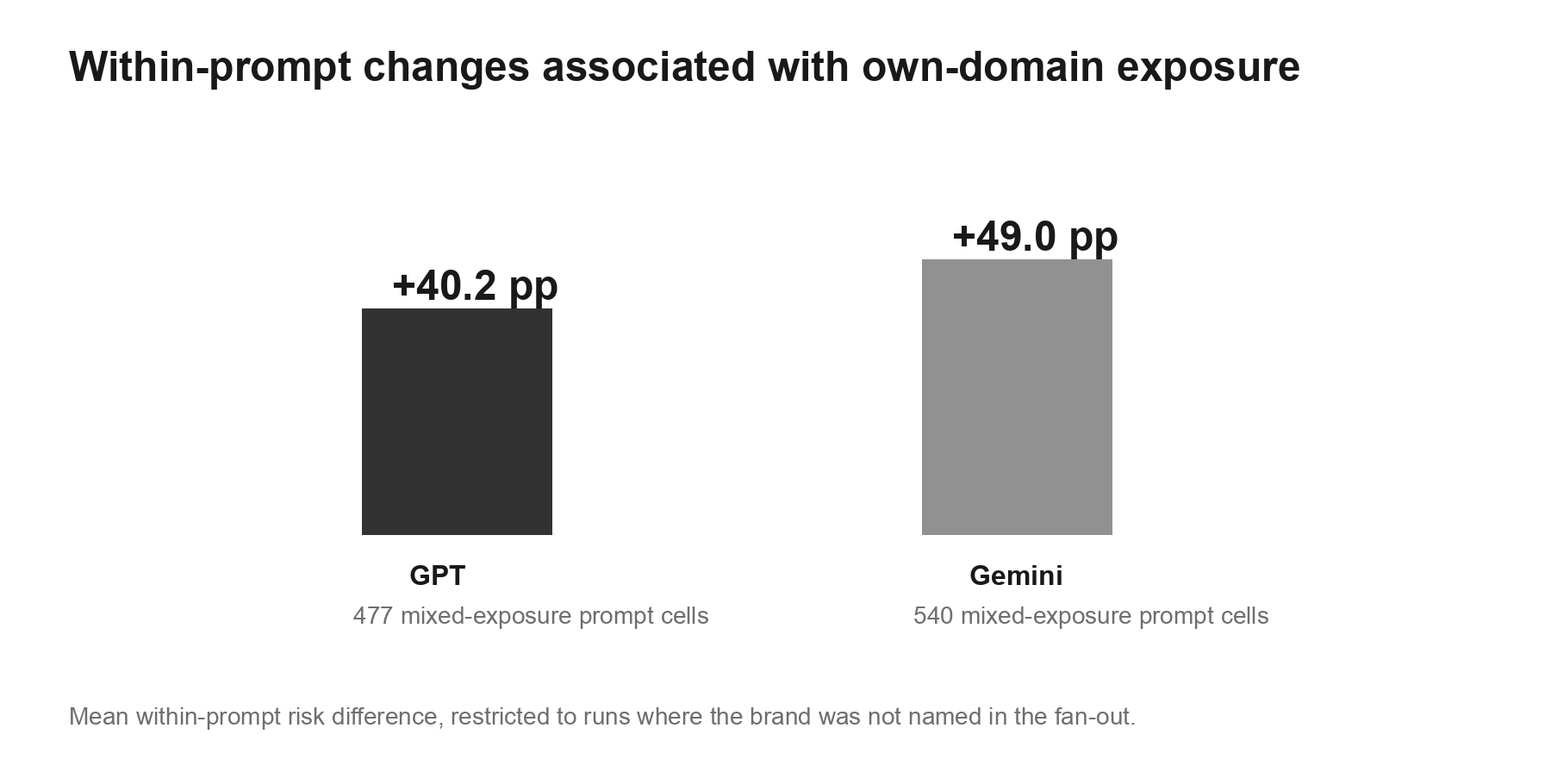}
\caption{Mean within-prompt risk difference associated with own-domain exposure, restricted to runs without branded fan-out.}
\label{fig:withinlarge}
\end{figure}

A Cochran--Mantel--Haenszel common odds ratio stratified by organization and prompt is 15.3 for GPT across 4,650 runs in exposure-varying strata and 29.7 for Gemini across 6,197 runs. Because each stratum holds the organization and prompt fixed, these estimates are not explained by time-invariant differences in brand identity or prompt wording. They remain observational: temporal changes in engine state, content, or third-party evidence can still confound the association.

Branded fan-out also varies within repeated cells, although less often. Its stratified common odds ratio is 8.6 on GPT across 81 varying strata and 58.9 on Gemini across 170 varying strata. The small number of varying GPT strata and sparse cells make these estimates less stable than the own-domain exposure result.

\subsection{The prior term is empirically measurable}

The large panel contains repeated measurements over time, which lets us estimate a brand-prompt prior without inspecting the current answer. For each observation after the first, define
\[
\hat P_{0,t}=\frac{m_{<t}+0.1}{n_{<t}+2},
\]
where $m_{<t}$ is the number of prior target-brand mentions and $n_{<t}$ the number of prior runs for the same organization-prompt-engine cell. The small pseudo-count shrinks early histories toward a low base rate.

\begin{figure}[H]
\centering
\includegraphics[width=0.94\linewidth]{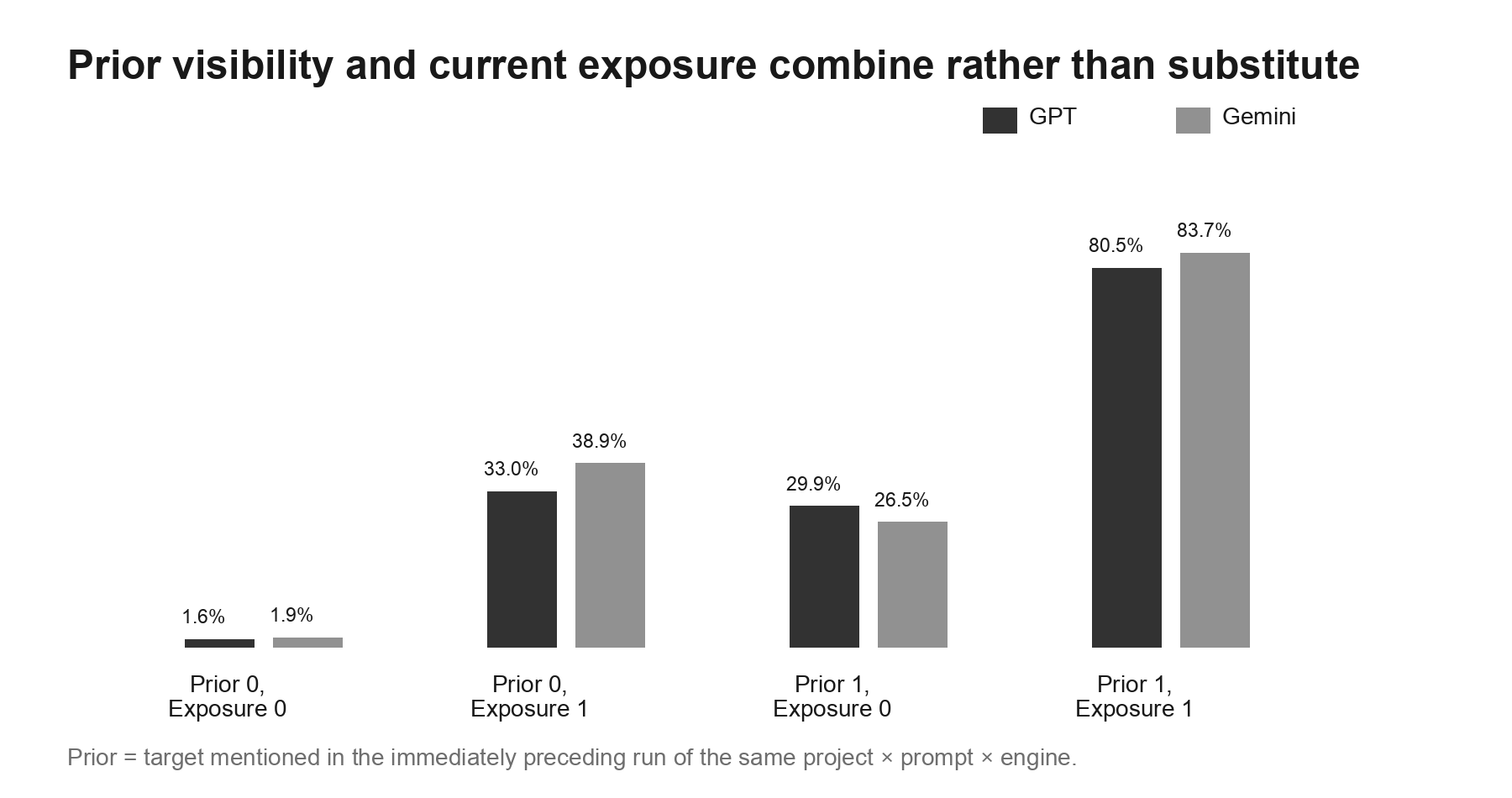}
\caption{Previous-run visibility and current own-domain exposure jointly stratify next-run brand mention.}
\label{fig:priorlarge}
\end{figure}

This prior is highly predictive. In low-prior GPT states (historical propensity below 0.1), brand mention is only 0.9\% when the own domain is not exposed and 25.1\% when it is exposed. In high-prior states (historical propensity at least 0.5), the corresponding rates are 32.0\% and 82.0\%. Gemini shows the same structure: 1.0\% versus 23.8\% in low-prior states and 28.7\% versus 86.2\% in high-prior states.

Thus live evidence and prior are complementary rather than interchangeable.

\subsection{Held-out predictive model}

We fit regularized logistic models on the earliest 70\% of sequential observations and evaluate on the latest 30\%, yielding 10,224 training and 4,383 held-out observations per engine. The \emph{prior-only} model uses the logit of historical propensity. The \emph{live-only} model uses own-domain exposure, branded fan-out, and Aiso prompt-intent labels. The \emph{full} model combines both. Model coefficients are fitted only on the training partition. During sequential evaluation, the history is updated after each completed observation, including earlier observations in the test period. This is a rolling diagnostic evaluation on monitored requests, not a fixed-origin forecast for unseen prompts. Current source citations and fan-outs are contemporaneous signals, not information available before the run.

\begin{table}[H]
\centering
\caption{Chronological held-out performance. Lower Brier score and log loss are better.}
\label{tab:heldout}
\begin{tabular}{llrrr}
\toprule
Engine & Model & AUC & Brier & Log loss \\
\midrule
GPT & Prior only & 0.937 & 0.0284 & 0.1034 \\
GPT & Live evidence only & 0.880 & 0.0318 & 0.1278 \\
GPT & Full & \textbf{0.963} & \textbf{0.0240} & \textbf{0.0871} \\
\addlinespace
Gemini & Prior only & 0.917 & 0.0699 & 0.2240 \\
Gemini & Live evidence only & 0.840 & 0.0824 & 0.2769 \\
Gemini & Full & \textbf{0.942} & \textbf{0.0610} & \textbf{0.1942} \\
\bottomrule
\end{tabular}
\end{table}

\begin{figure}[H]
\centering
\includegraphics[width=0.92\linewidth]{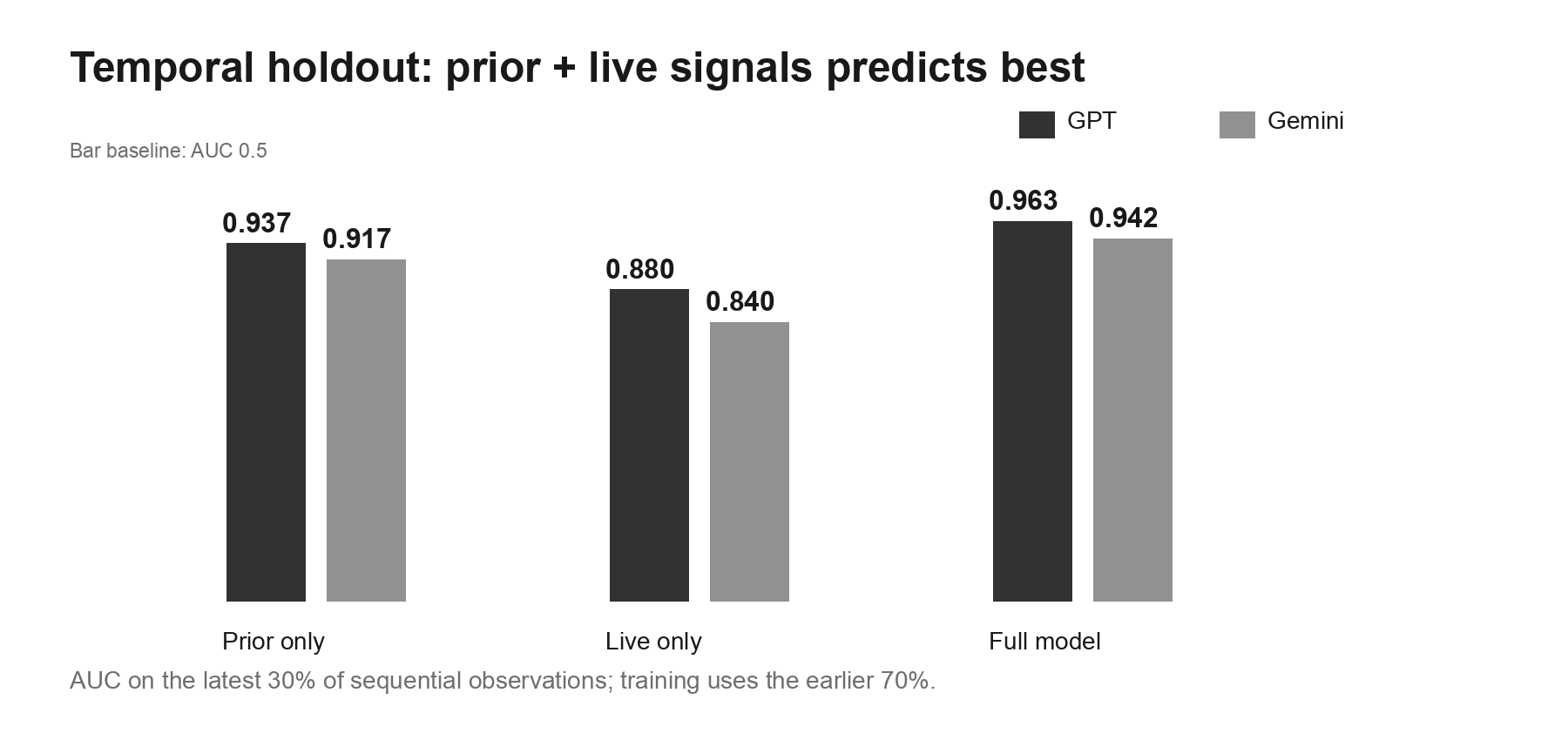}
\caption{Held-out AUC for historical prior, live evidence, and their combination.}
\label{fig:heldout}
\end{figure}

Adding live evidence to the historical prior reduces held-out Brier error by 15.4\% on GPT and 12.8\% on Gemini. In the full logistic model, an own-domain citation has an estimated odds ratio of 14.7 on GPT and 20.5 on Gemini after controlling for historical prior and intent. Branded fan-out has estimated odds ratios of 3.5 and 22.5 respectively. These coefficients describe this panel and should not be interpreted as invariant engine constants.

The model also exposes cross-engine divergence. For the same snapshot-prompt observations, GPT and Gemini agree on target-brand mention 84.7\% of the time, but 1,999 paired observations are Gemini-only mentions and 680 are GPT-only mentions. Own-domain exposure agrees 81.1\% of the time. This is consistent with the cross-engine source divergence documented in our companion paper.

\subsection{Manual-prompt sensitivity}

Prompt provenance could otherwise be a concern because some monitoring prompts are generated by Aiso's enrichment pipeline. We therefore repeat the analysis on the manually curated subset only. It contains 1,569 distinct unbranded prompts and 27,000 engine observations before requiring prior history. The basic exposure result remains large: own-domain exposure without branded fan-out is associated with a 41.5\% GPT mention rate versus 2.6\% without either live signal, and 57.3\% versus 3.0\% on Gemini.

The temporal models are also stable. Among 11,931 sequential observations per engine, the manually curated full model reaches held-out AUC 0.960 on GPT and 0.943 on Gemini, compared with 0.932 and 0.918 for prior history alone. This sensitivity makes the core conclusion independent of generated monitoring prompts.

\subsection{Model-version sensitivity}

The GPT panel spans a model transition. On GPT-5 nano, the no-live-signal mention rate is 1.8\% and own-domain-only rate is 45.1\%. On GPT-5.4, the corresponding rates are 6.3\% and 56.3\%. The absolute baseline therefore changes across model versions, but the exposure gradient remains. Gemini 2.5 Flash shows 3.8\% and 58.4\%. This is another reason not to treat any coefficient as a permanent engine constant.

\section{A Practical AI Visibility Equation}

\subsection{Why the mnemonic is useful}

The large-panel results support a more precise probabilistic decomposition. Let $P_0$ denote the historical or citation-free prior probability that the brand is selected, $L$ a live-evidence event generated by match and engine retrieval, and $S$ final selection conditional on live evidence. A useful noisy-OR-style approximation is

\begin{equation}
P(M=1) \approx P_0 + (1-P_0)P(L=1)P(S=1\mid L=1).
\label{eq:fitted}
\end{equation}

For optimization, $P(L=1)$ can itself be decomposed conceptually into request/page match and engine exposure:
\[
P(L=1) \approx f(\text{Match},\text{fan-out},\text{retrieval},e,t).
\]
The practical mnemonic is therefore \emph{visibility $\approx$ prior $+$ $(1-prior)\times match\times exposure\times selection$}. The fitted logistic model operationalizes the prior and observable live-evidence terms without pretending that the multiplicative mnemonic is the engine's hidden ranking algorithm.

The multiplication sign captures a simple operational truth. On the live-retrieval path, a weak upstream stage can bottleneck later strengths. Excellent selection propensity is irrelevant if the source is never exposed. Exposure is wasted if the available page does not answer the request. A perfect page can still be invisible if the engine's fan-out never retrieves it.

The additive prior term prevents a second error: assuming that all recommendations are caused by live search. The Organization B benchmark shows that some brands have large citation-free mention volume.

\subsection{Operational definitions}

For an AI-visibility system, each term can be measured separately.

\paragraph{Match.}
Given a real request $R$, score the best available owned page and relevant third-party pages. The output is not "AI readiness." It is prompt-to-page fit. BM25, embeddings, cross-encoders, or calibrated hybrids can all be used, provided their estimand is explicit.

\paragraph{Exposure.}
Measure whether the engine searches, which fan-out queries it issues, which domains rank or are retrieved, and which sources are exposed or cited. This is engine- and time-specific.

\paragraph{Selection.}
Condition on observable exposure and measure whether the brand is named, recommended, compared, ranked, or used as supporting evidence.

\paragraph{Prior.}
Measure brand appearance when visible live evidence is absent or insufficient to explain the answer. This can be estimated descriptively from citation-free runs, no-search controls, or carefully designed memory-versus-search experiments. It should not automatically be called "training data" because the underlying mechanism may be unobservable.

\subsection{An SEO analogy with an important difference}

A rough SEO analogy would be:

\begin{center}
\textbf{Traditional search: relevance + authority.}\\
\textbf{Generative search: match + exposure + selection + prior.}
\end{center}

The important difference is that generative search inserts an answer-composition layer after retrieval. Traditional ranking decides what result is shown and where. Generative systems can retrieve a page and then omit the brand, cite a page without using it strongly, or mention a brand without citing its own site.

This makes attribution harder but diagnosis richer.

\section{Implications for Measurement and Optimization}

\subsection{Use real demand before optimizing pages}

The first measurement question should not be "is this page optimized for AI?" It should be "which real requests is this page supposed to satisfy?" Our prior work shows that conversation state can contain constraints absent from the final prompt \citep{tannenbaum2026prompt,tannenbaum2026context}. The current fan-out data adds another transformation: even after the human request is known, the engine may search for a different or expanded representation.

A practical workflow therefore begins with real conversations or real prompt logs, not a synthetic keyword list alone.

\subsection{Treat fan-outs as hidden query demand}

The real-prompt cohort shows that commercial requests disproportionately trigger fan-out. This means fan-out analysis is not just a technical curiosity. It identifies the evaluation criteria that a generative engine itself decides to investigate.

For the mascara example, the user asks for a new mascara. The system searches both "what to look for" and "best mascaras." A brand that only has a product page may match the second query but not the first. A guide that explains brush shape, formula, ingredients, wear, and use case can cover the evaluation query while product pages cover the shortlist query.

The same pattern appears in software. "Suggest one accounting tool" fans out into both a best-tool query and a feature/criteria query. The hidden demand surface is broader than the head prompt.

\subsection{Separate content work from distribution work}

The Organization A analysis suggests two different optimization problems.

If page match is low, the likely intervention is content-side: create or improve a page that answers the request directly.

If page match is high but exposure is low, more content on the same domain may not solve the bottleneck. The intervention may be distribution-side: indexing, crawlability, third-party coverage, comparison pages, directories, partnerships, source authority, or whatever increases the probability that the engine's retrieval layer sees brand-supporting evidence.

This is the generative-search analogue of distinguishing relevance work from authority/distribution work in SEO.

\subsection{Measure each engine separately}

The same BM25 best-page match has AUC 0.641 for Gemini citation and 0.545 for GPT citation. This is a material difference in this cohort. Our companion paper finds that engines also expose largely different URL sets for the same prompt \citep{tannenbaum2026engine}.

Therefore a single engine-agnostic "visibility probability" requires strong assumptions. A safer interface reports page fit once, then exposure and selection separately by engine.

\subsection{Do not confuse citation with recommendation}

The Organization B benchmark shows that citation and recommendation can diverge dramatically. Competitor 1 and Competitor 2 appear many times without own-domain citations. In practical monitoring, a cited-source dashboard and a brand-mention dashboard answer different questions.

A third-party page can also carry the brand into the answer. This is why source mapping should include pages \emph{about} the brand, not only pages \emph{owned} by the brand.

\section{Relationship to the Previous Papers}

The six-paper sequence now forms a coherent measurement stack.

\begin{enumerate}[leftmargin=*]
\item \textbf{Answer-Reconstruction Search Density} asks how much conventional search work can be compressed into one AI answer \citep{tannenbaum2026arsd}.
\item \textbf{The Prompt Is Not the Query} shows that the final visible prompt is often an incomplete representation of the active request state \citep{tannenbaum2026prompt}.
\item \textbf{Beyond the Final Prompt} experimentally shows that restoring the missing conversation context changes answers \citep{tannenbaum2026context}.
\item \textbf{Purchase Advice and Observable Buyer Responses} separates visible recommendation behavior from later commercial outcomes \citep{tannenbaum2026purchase}.
\item \textbf{Scoring With the Engine} separates page-side scoring from engine-specific source exposure \citep{tannenbaum2026engine}.
\item \textbf{This paper} connects those layers into a stage model and adds real-prompt fan-out, page-match, evidence-exposure, selection, and citation-free-prior measurements.
\end{enumerate}

The sequence matters because each paper removes one shortcut. The prompt is not necessarily the request. The answer is not a single search. Page quality is not exposure. Citation is not recommendation. Recommendation is not purchase.

\section{Robustness and Alternative Explanations}

\subsection{Could own-domain citation merely be a consequence of brand mention?}

The Organization A export is observational. The stored answer and sources are co-produced by the engine, so temporal ordering inside the system is not observed. We therefore do not claim that inserting an Organization A URL into a source list would causally increase mentions tenfold.

However, the association is still diagnostically important. A system that predicts brand visibility without measuring source exposure is omitting a variable that almost perfectly partitions GPT mentions and strongly partitions Gemini mentions in this cohort.

\subsection{Could match be underestimated by lexical scoring?}

Yes. BM25 is intentionally conservative. It can miss semantic equivalence, paraphrase, location aliases, and entity relationships. A cross-encoder or embedding model could produce stronger page-fit discrimination, consistent with \citet{bajemon2026}. We use lexical scoring because it is transparent, zero-cost to reproduce, and sufficient to demonstrate that even a useful match signal is not the same as exposure.

The engine asymmetry is especially important here. If the same lexical metric predicts Gemini better than GPT, either the engines weight relevance differently, their retrieval systems expose different candidates, or the measurement noise differs. The present data do not distinguish those explanations.

\subsection{Could the prior-compatible path be third-party retrieval rather than model memory?}

Yes. "Prior" is a mnemonic, not a mechanistic claim. A brand mentioned without its own-domain citation may enter through third-party pages, internal model knowledge, latent retrieval, prior conversation context, or other evidence not visible in the stored citation list.

The Organization B citation-free prompts are stronger because no citations are produced at all, but even there we cannot prove that no retrieval occurred internally. We therefore use "prior-compatible" throughout.

\subsection{Could the Organization A site have changed after the July visibility run?}

Yes. The page crawl is from 19 September 2026, later than the 6 July visibility export. Organization A added and updated content during that period. The match score is therefore a current coverage measure applied retrospectively to older engine outcomes. This may overstate historical fit for prompts whose supporting content was added later.

This limitation makes the positive Gemini match relationship more interesting but less causal. A future replication should freeze page snapshots at the same time as engine runs.

\section{Limitations}
\label{sec:limits}

Several limitations bound interpretation.

First, the datasets were collected for operational research rather than designed as one preregistered experiment. The paper deliberately treats them as a triangulation across stages, not as a single causal panel.

Second, the real-prompt fan-out cohort is small. Only 20 of 80 unique prompts trigger observed fan-out, and the intent mix is not population-representative. The 21.5-fold commercial-versus-informational risk ratio is therefore a cohort result, not an estimate of ChatGPT behavior for all users.

Third, fan-out visibility depends on instrumentation. A prompt with no recorded \texttt{search\_query} is treated as not triggering an observed fan-out. Hidden retrieval actions would weaken that interpretation.

Fourth, the Organization A page-match crawl is temporally later than the engine visibility export. The analysis should be read as a retrospective coverage association.

Fifth, own-domain citation is an incomplete measure of evidence exposure. Third-party pages may mention the brand and can be more important than owned pages. This likely explains some no-own-citation brand mentions.

Sixth, brand mention is not equivalent to a positive recommendation, rank, click, lead, or purchase. Our fourth paper shows how quickly observability falls after recommendation \citep{tannenbaum2026purchase}.

Seventh, the datasets use a limited set of verticals and engines. The effects should be replicated on larger, prospectively frozen cohorts.

Finally, all model and search behavior is time-dependent. The paper is a dated measurement of systems observed in 2026, not a claim about permanent engine internals.

\section{Future Work}

The stage model suggests a more rigorous next experiment.

For each real user prompt, freeze:
\begin{enumerate}[leftmargin=*]
\item the full request state;
\item the target brand's page corpus;
\item the engine and model version;
\item every observable fan-out/search query;
\item all retrieved/ranked candidate URLs;
\item the final cited source set;
\item the final answer and brand position.
\end{enumerate}

Then intervene on one stage at a time. Improve page fit without changing distribution. Add a third-party source without changing the owned page. Improve crawlability without changing copy. Run no-search and forced-search conditions. Repeat enough times to estimate stochastic variance.

Such a design could turn the mnemonic in Equation~\ref{eq:mnemonic} into a genuinely calibrated stage model.

A second direction is to make Match conversation-aware. Our prior papers show that a standalone prompt can omit earlier budget, geography, preference, or product constraints. Prompt-to-page fit should ultimately be computed against the active request state, not only the final user string.

A third direction is to distinguish \emph{evidence exposure} from \emph{citation}. The most useful future source graph would mark whether the brand appears inside every retrieved third-party page, whether that page is cited, and whether the cited passage actually supports the brand mention.

\section{Conclusion}

There is a useful ranking-factor insight in recent GEO research, but it is narrower than "make the page better and AI will cite it."

Query-page match matters. In our Organization A cohort it meaningfully predicts Gemini source exposure and final mention. But match is only one stage.

Real user prompts are often rewritten into commercial fan-outs before retrieval. Engines then decide which evidence to expose. Evidence exposure is associated with roughly an order-of-magnitude difference in brand-mention probability in both GPT and Gemini. Finally, established brands can appear without their own domain being cited, creating a second prior-compatible path that page scoring and citation tracking alone cannot explain.

The resulting practical model is:

\begin{equation*}
\boxed{P(M=1) \approx P_0+(1-P_0)P(L=1)P(S=1\mid L=1).}
\end{equation*}

In practitioner language, visibility is prior plus the remaining opportunity carried through match, exposure, and selection.

The formula is useful precisely because it is not one score. It tells a marketer what to diagnose next.

The large-panel validation sharpens this diagnostic: historical prior and live evidence are complementary, and their combination outperforms either alone on a chronological holdout. If Match is weak, improve the page. If Match is strong but Exposure is weak, work on retrieval and distribution. If Exposure is strong but Selection is weak, study how the answer chooses among available evidence. If the Prior path is weak, the longer-run brand footprint remains a separate problem.

That is a closer analogue to the relevance-and-authority logic of SEO than a generic "AI optimization score." It starts with real demand and keeps the engine in the measurement loop.

\section*{Data Governance and Conflict of Interest}

The empirical material is described in this paper as Aiso proprietary research data. Aiso performs substantial processing, deduplication, normalization, intent and brand enrichment, provenance tracking, prompt construction/curation, source extraction, and longitudinal monitoring before the analytic tables used here are produced. No raw source corpus is redistributed with this submission.

All organizations in the empirical sections are anonymized. Public artifacts contain only aggregate results, reproducibility code that operates on aggregate inputs, and short non-identifying prompt fragments. They exclude raw conversations, complete prompts, client names, project identifiers, account information, source exports, and private collection names.

The author is the founder and CEO of Aiso Boost Ltd., which develops commercial AI-search measurement software. This creates a financial conflict of interest. We therefore report weak and null findings, distinguish predictive associations from causal effects, preserve engine-specific results, and make the fitted equation explicitly diagnostic rather than claiming knowledge of proprietary ranking algorithms.

\appendix

\section{Reproducibility Notes}

The submission includes:
\begin{itemize}[leftmargin=*]
\item \texttt{anc/aggregate\_results.csv}, containing the principal aggregate estimates;
\item \texttt{anc/make\_figures.py}, generating the manuscript figures from aggregate values, with accompanying aggregate model outputs;
\item the normalized BM25 scoring specification in Equation~\ref{eq:bm25};
\item counts and model summaries supporting the reported descriptive proportions and primary risk ratios.
\end{itemize}

The proprietary raw Aiso conversation corpus and client-level monitoring records are not included because privacy and commercial data-governance constraints do not permit redistribution. The released aggregates reproduce figures and summary arithmetic, but do not permit an independent refit of the models, reconstruction of bootstrap intervals, or reconstruction of page-match scores.

\section{Key Counts}

\begin{table}[H]
\centering
\caption{Counts underlying the primary results.}
\begin{tabular}{lrrr}
\toprule
Comparison & Total & Positive & Rate \\
\midrule
Real commercial prompts triggering fan-out & 23 & 18 & 78.3\% \\
Real informational prompts triggering fan-out & 55 & 2 & 3.6\% \\
GPT rows with own-domain citation and brand mention & 30 & 30 & 100.0\% \\
GPT rows without own-domain citation and brand mention & 169 & 16 & 9.5\% \\
Gemini rows with own-domain citation and brand mention & 65 & 48 & 73.8\% \\
Gemini rows without own-domain citation and brand mention & 134 & 11 & 8.2\% \\
Organization B citation-free runs mentioning Competitor 1 & 40 & 30 & 75.0\% \\
Organization B citation-free runs mentioning Competitor 2 & 40 & 28 & 70.0\% \\
Organization B citation-free runs mentioning Organization B & 40 & 0 & 0.0\% \\
\bottomrule
\end{tabular}
\end{table}

\section{Effect Summary}

\begin{table}[H]
\centering
\caption{Primary Organization A effect estimates.}
\begin{tabular}{llrr}
\toprule
Outcome & Engine & Estimate & 95\% bootstrap CI \\
\midrule
Match $\rightarrow$ own citation, AUC & GPT & 0.545 & 0.429--0.658 \\
Match $\rightarrow$ own citation, AUC & Gemini & 0.641 & 0.556--0.722 \\
Match $\rightarrow$ brand mention, AUC & GPT & 0.539 & 0.443--0.637 \\
Match $\rightarrow$ brand mention, AUC & Gemini & 0.608 & 0.519--0.695 \\
Own citation $\rightarrow$ mention, RR & GPT & 10.56 & 7.04--18.89 \\
Own citation $\rightarrow$ mention, RR & Gemini & 9.00 & 5.47--18.88 \\
Q4/Q1 match $\rightarrow$ citation, RR & Gemini & 2.27 & 1.32--4.74 \\
Q4/Q1 match $\rightarrow$ mention, RR & Gemini & 2.09 & 1.20--4.29 \\
\bottomrule
\end{tabular}
\end{table}


\begin{thebibliography}{99}

\bibitem[Aggarwal et~al.(2024)]{aggarwal2024}
Pranjal Aggarwal, Vishvak Murahari, Tanmay Rajpurohit, Ashwin Kalyan, Karthik Narasimhan, and Ameet Deshpande.
\newblock GEO: Generative Engine Optimization.
\newblock \emph{arXiv preprint arXiv:2311.09735}, 2023; KDD, 2024.

\bibitem[Allaham and Diakopoulos(2026)]{allaham2026}
Mowafak Allaham and Nicholas Diakopoulos.
\newblock Synthetic Sources?: Auditing Generative Search Engine Citations for Evidence of AI-Generated Sources.
\newblock \emph{arXiv preprint arXiv:2605.23684}, 2026.

\bibitem[Asai et~al.(2024)]{asai2024}
Akari Asai, Zeqiu Wu, Yizhong Wang, Avirup Sil, and Hannaneh Hajishirzi.
\newblock Self-RAG: Learning to Retrieve, Generate, and Critique through Self-Reflection.
\newblock In \emph{ICLR}, 2024.

\bibitem[Bajemon and Rochet(2026)]{bajemon2026}
Elisha Bajemon and Andre-Louis Rochet.
\newblock Scoring Without the Engine: Validating a Deterministic, Manipulation-Resistant Content Score for Generative Engines, End to End.
\newblock \emph{arXiv preprint arXiv:2609.07559}, 2026.

\bibitem[Brin and Page(1998)]{brin1998}
Sergey Brin and Lawrence Page.
\newblock The Anatomy of a Large-Scale Hypertextual Web Search Engine.
\newblock \emph{Computer Networks and ISDN Systems}, 30(1--7):107--117, 1998.

\bibitem[Gao et~al.(2023)]{gao2023}
Tianyu Gao, Howard Yen, Jiatong Yu, and Danqi Chen.
\newblock Enabling Large Language Models to Generate Text with Citations.
\newblock In \emph{Proceedings of EMNLP}, 2023.

\bibitem[Lewis et~al.(2020)]{lewis2020}
Patrick Lewis, Ethan Perez, Aleksandra Piktus, Fabio Petroni, Vladimir Karpukhin, Naman Goyal, Heinrich K{\"u}ttler, Mike Lewis, Wen-tau Yih, Tim Rockt{\"a}schel, Sebastian Riedel, and Douwe Kiela.
\newblock Retrieval-Augmented Generation for Knowledge-Intensive NLP Tasks.
\newblock In \emph{NeurIPS}, 2020.

\bibitem[Lin et~al.(2021)]{lin2021}
Sheng-Chieh Lin, Jheng-Hong Yang, and Jimmy Lin.
\newblock Contextualized Query Embeddings for Conversational Search.
\newblock In \emph{Proceedings of EMNLP}, 2021.

\bibitem[Qian and Dou(2022)]{qian2022}
Hongjin Qian and Zhicheng Dou.
\newblock Explicit Query Rewriting for Conversational Dense Retrieval.
\newblock In \emph{Proceedings of EMNLP}, pages 4725--4737, 2022.
\newblock DOI: 10.18653/v1/2022.emnlp-main.311.

\bibitem[Robertson and Zaragoza(2009)]{robertson2009}
Stephen Robertson and Hugo Zaragoza.
\newblock The Probabilistic Relevance Framework: BM25 and Beyond.
\newblock \emph{Foundations and Trends in Information Retrieval}, 3(4):333--389, 2009.


\bibitem[Chen et~al.(2025)]{chen2025geo}
Mahe Chen, Xiaoxuan Wang, Kaiwen Chen, and Nick Koudas.
\newblock Generative Engine Optimization: How to Dominate AI Search.
\newblock \emph{arXiv preprint arXiv:2509.08919}, 2025.

\bibitem[Grossman et~al.(2026)]{grossman2026}
Riley Grossman, Songjiang Liu, Michael K. Chen, Mike Smith, Cristian Borcea, and Yi Chen.
\newblock How Generative AI Disrupts Search: An Empirical Study of Google Search, Gemini, and AI Overviews.
\newblock \emph{arXiv preprint arXiv:2604.27790}, 2026.

\bibitem[Lee et~al.(2025)]{lee2025setr}
Dahyun Lee, Yongrae Jo, Haeju Park, and Moontae Lee.
\newblock Shifting from Ranking to Set Selection for Retrieval Augmented Generation.
\newblock In \emph{Proceedings of ACL}, pages 17606--17619, 2025. DOI: 10.18653/v1/2025.acl-long.861.

\bibitem[Li et~al.(2024)]{li2024dmqr}
Zhicong Li, Jiahao Wang, Zhishu Jiang, Hangyu Mao, Zhongxia Chen, Jiazhen Du, Yuanxing Zhang, Fuzheng Zhang, Di Zhang, and Yong Liu.
\newblock DMQR-RAG: Diverse Multi-Query Rewriting for RAG.
\newblock \emph{arXiv preprint arXiv:2411.13154}, 2024.

\bibitem[Ma et~al.(2023)]{ma2023rewrite}
Xinbei Ma, Yeyun Gong, Pengcheng He, Hai Zhao, and Nan Duan.
\newblock Query Rewriting in Retrieval-Augmented Large Language Models.
\newblock In \emph{Proceedings of EMNLP}, pages 5303--5315, 2023. DOI: 10.18653/v1/2023.emnlp-main.322.

\bibitem[Martinez(2026)]{martinez2026}
Olivier Martinez.
\newblock Optimizing Visibility in Generative Engines: A Critical Survey of Generative Engine Optimization (2023--2026).
\newblock \emph{arXiv preprint arXiv:2607.14035}, 2026.

\bibitem[Medrano et~al.(2026)]{medrano2026}
Luigi Medrano, Arush Verma, and Mukul Chhabra.
\newblock Scaling Retrieval Augmented Generation with RAG Fusion: Lessons from an Industry Deployment.
\newblock \emph{arXiv preprint arXiv:2603.02153}, 2026.

\bibitem[Mo et~al.(2023)]{mo2023}
Fengran Mo, Kelong Mao, Yutao Zhu, Yihong Wu, Kaiyu Huang, and Jian-Yun Nie.
\newblock ConvGQR: Generative Query Reformulation for Conversational Search.
\newblock In \emph{Proceedings of ACL}, pages 4998--5012, 2023. DOI: 10.18653/v1/2023.acl-long.274.

\bibitem[Mody(2026)]{mody2026}
Deepanshu Mody.
\newblock The Attribution-Compression Frontier in Retrieval-Augmented Generation.
\newblock \emph{arXiv preprint arXiv:2609.14245}, 2026.

\bibitem[Nematov et~al.(2025)]{nematov2025}
Ikhtiyor Nematov, Tarik Kalai, Elizaveta Kuzmenko, Gabriele Fugagnoli, Dimitris Sacharidis, Katja Hose, and Tomer Sagi.
\newblock Source Attribution in Retrieval-Augmented Generation.
\newblock \emph{arXiv preprint arXiv:2507.04480}, 2025.

\bibitem[Rackauckas(2024)]{rackauckas2024}
Zackary Rackauckas.
\newblock RAG-Fusion: a New Take on Retrieval-Augmented Generation.
\newblock \emph{arXiv preprint arXiv:2402.03367}, 2024.

\bibitem[Strauss et~al.(2025)]{strauss2025}
Ilan Strauss, Jangho Yang, Tim O'Reilly, Sruly Rosenblat, and Isobel Moure.
\newblock The Attribution Crisis in LLM Search Results.
\newblock \emph{arXiv preprint arXiv:2508.00838}, 2025.

\bibitem[Ye et~al.(2023)]{ye2023}
Fanghua Ye, Meng Fang, Shenghui Li, and Emine Yilmaz.
\newblock Enhancing Conversational Search: Large Language Model-Aided Informative Query Rewriting.
\newblock In \emph{Findings of EMNLP}, pages 5985--6006, 2023. DOI: 10.18653/v1/2023.findings-emnlp.398.

\bibitem[Tannenbaum(2026a)]{tannenbaum2026arsd}
Benjamin Tannenbaum.
\newblock Answer-Reconstruction Search Density: Measuring the Query and Source Work Compressed by Conversational Answers.
\newblock \emph{arXiv preprint arXiv:2607.18904}, 2026.

\bibitem[Tannenbaum(2026b)]{tannenbaum2026prompt}
Benjamin Tannenbaum.
\newblock The Prompt Is Not the Query: How Request State Evolves Across Multi-Turn AI Conversations.
\newblock \emph{arXiv preprint arXiv:2607.22392}, 2026.

\bibitem[Tannenbaum(2026c)]{tannenbaum2026context}
Benjamin Tannenbaum.
\newblock Beyond the Final Prompt: Measuring the Effect of Within-Conversation Context on AI Answers.
\newblock \emph{arXiv preprint arXiv:2608.02556}, 2026.

\bibitem[Tannenbaum(2026d)]{tannenbaum2026purchase}
Benjamin Tannenbaum.
\newblock Purchase Advice and Observable Buyer Responses in Real AI Conversations.
\newblock \emph{arXiv preprint arXiv:2609.09878}, 2026.

\bibitem[Tannenbaum(2026e)]{tannenbaum2026engine}
Benjamin Tannenbaum.
\newblock Scoring With the Engine: Retrieval Exposure, Cross-Engine Divergence, and the Limits of Engine-Agnostic GEO Scores.
\newblock Submitted manuscript, 2026.
\newblock \url{https://www.getaiso.com/papers/scoring-with-the-engine-2026.pdf}.

\bibitem[Wu et~al.(2022)]{wu2022}
Zeqiu Wu, Yi Luan, Hannah Rashkin, David Reitter, Hannaneh Hajishirzi, Mari Ostendorf, and Gaurav Singh Tomar.
\newblock CONQRR: Conversational Query Rewriting for Retrieval with Reinforcement Learning.
\newblock In \emph{Proceedings of EMNLP}, pages 10000--10014, 2022.
\newblock DOI: 10.18653/v1/2022.emnlp-main.679.

\bibitem[Yang(2025)]{yang2025}
Kai-Cheng Yang.
\newblock News Source Citing Patterns in AI Search Systems.
\newblock \emph{arXiv preprint arXiv:2507.05301}, 2025.

\end{thebibliography}
\end{document}